\documentclass[conference]{IEEEtran}

\usepackage[english]{babel}
\usepackage[utf8x]{inputenc}
\usepackage[T1]{fontenc}
\usepackage{listings}
\lstdefinelanguage{solidity}
{morekeywords={function, struct, if, else, return, event, delete, contract, public, returns, modifier, send, indexed},
sensitive=false,
morecomment=[l]{//},
morecomment=[s]{/*}{*/},
morestring=[b]",
}

\usepackage[a4paper,top=3cm,bottom=2cm,left=3cm,right=3cm,marginparwidth=1.75cm]{geometry}

\usepackage{amsmath}
\usepackage{graphicx}
\usepackage[colorinlistoftodos]{todonotes}

\title{Smart Contract SLAs for Dense Small-Cell-as-a-Service}
\author{Emanuele Di Pascale, Jasmina McMenamy, Irene Macaluso, Linda Doyle
\thanks{The authors are with CONNECT centre, Trinity College Dublin.}}

\begin{document}
\lstset{language=solidity}
\maketitle

\begin{abstract}
The disruptive power of blockchain technologies represents a great opportunity to re-imagine standard practices of providing radio access services  by addressing critical areas such as deployment models that can benefit from brand new approaches. As a starting point for this debate, we look at the current limits of infrastructure sharing, and specifically at the Small-Cell-as-a-Service trend, asking ourselves how we could push it to its natural extreme: a scenario in which any individual home or business user can become a service provider for mobile network operators (MNOs), freed from all the scalability and legal constraints that are inherent to the current modus operandi. We propose the adoption of smart contracts to implement simple but effective Service Level Agreements (SLAs) between small cell providers and MNOs, and present an example contract template based on the Ethereum blockchain.
\end{abstract}

\vspace{-7pt}
\section{Introduction}
\label{sec:introduction}
\vspace{-6pt}
The ownership model of mobile network infrastructure has evolved from a monolithic to a composite structure. Since the early 2000s, passive sharing has started to gain a lot of attention from the industry; nowadays, the spin-off or sale of tower holdings by mobile network operators (MNOs) to tower companies is common practice. Active sharing, which involves the sharing of Radio Access Network (RAN) nodes, the backhaul or the core network, is also becoming more and more important, and RAN sharing agreements can be found in many European countries. Furthermore, we observe an increasing interest in Small-Cell-as-a-Service (SCaaS), in which a third-party provisions radio access capacity to  MNOs in localized areas \cite{ RealWireless}.  

The solutions for SCaaS currently consider medium to large scale venues (e.g. shopping malls, hospitals, etc.) \cite{scf2016}. Agreements between MNOs and Small Cell Providers (SCPs) are negotiated between the two parties and lead to long-term arrangements. In \cite{scf2016}, an analysis of the SCaaS model connects the slow progress of deployment to the uncertainty over the business models. While network densification is  recognized as a crucial part of future networks, it seems unlikely that small to very-small SCPs will be able to negotiate deals with MNOs. In other words,  network ownership diversification seems to have reached its limit. However, is this necessarily the case? In this paper we investigate Smart Contracts as possible enablers for advancing the trajectory of mobile network ownership.

The term “smart contract” does not have a universally accepted definition, but in the context of this paper it refers to a block of code registered on the ledger of a blockchain platform, which is able to automatically enforce contractual clauses between agreeing parties. 
As the source code for these contracts is publicly visible, and the blockchain platform on which the code resides guarantees immutability, actors interacting through a smart contract can rest assured that no spurious behavior will be possible from any of the contracting parties, thus creating the foundation for trusted interactions in a trustless environment. Furthermore, smart contracts can greatly reduce the costs associated to drafting a binding agreement compared to standard legal practices. Finally, the ability to quickly move funds through the cryptocurrency underlying the blockchain platform on which the smart contract resides can make these transactions cheaper to execute and faster to  settle. While there are still a number of issues surrounding the use of smart contracts, the numerous benefits highlighted above make for a very strong case for their adoption in all those scenarios where traditional legal agreements would be too costly or too cumbersome. 

\vspace{-7pt}
\section{Smart Contract SLAs}
\label{sec:smart_contracts}
\vspace{-5pt}
Before discussing how smart contracts can be introduced into SCaaS model, we highlight the features and requirements of the current cellular systems to support  SCaaS advocated in this paper, namely:
i) Small cells are required to support multi-tenancy with a number of MNOs which share RAN elements in a dynamic manner; ii) 3GPP LTE Multiple Operator Core Network feature already enables sharing in the RAN, supporting connections to multiple core networks using S1-flex interfaces; iii) Self-configuration, self-optimization and self-healing capabilities  that are a part of LTE Self-Organizing Networks feature set handling e.g. automatic setup and optimization of mobility need to be further investigated to operate in a multi-tenancy environment \cite{sallent2016multi}; iv) From a spectrum perspective, dedicated spectrum from MNOs may be used - requiring agreement with the SCP on how to manage the radio resources - or unlicensed spectrum could be employed using schemes such as LTE unlicensed in the 5 GHz band or using mmW spectrum. 

The Service Level Agreement (SLA) between an SCP and an MNO will invariably include specification of QoS parameters, where the so-called QoS Class Identifiers (QCI) are used to characterize  a service in terms of priority, packet delay budget and packet loss rate. Well-defined Key Performance Indicators (KPIs) will be then used to measure the performance, which will often be specified per QCI.

In the paper, we developed a sample smart contract implementing a basic SLA between an SCP and an MNO, using Ethereum \cite{ethereum} as the reference platform. The full code of the contract can be found online in the repository referenced in \cite{repo}. 
This example assumes that the SCP and the MNO have already agreed to enter a partnership on the basis of one of the template contracts developed by the MNO. This could be done through an ad-hoc online platform managed by the MNO through which providers that match certain criteria can apply, or possibly through a third-party asset register in which SCPs can register the available infrastructure for the benefit of interested MNOs. 

\emph{Events} are used to enable asynchronous notifications between the MNO and the SCPs. For instance, a \texttt{PeriodicPayout} event notifies the SCP that payment proportional to the traffic serviced is available and an \texttt{InsufficientThroughput} event is fired on detection of a breach of the agreed level of service with regards to throughput. Events are permanently stored in the blockchain, and can easily be retrieved e.g. by a Javascript client, which can be set to listen for specific event or to query events for any of their indexed parameters. We assume that payments are made periodically based on the amount of traffic served, as measured by the MNO, and that the price per kb of data was agreed as part of the agreement. The payment uses the withdrawal pattern, following best-practice principles of Ethereum contracts. The payment functions can easily be tweaked to accommodate different paying mechanisms, e.g., such as a periodic flat-rate payment for best-effort service; indeed it is expected that the MNO will have different SLA smart contracts for different type of SCPs.

Finally, we describe the procedure of infraction for breaching the
agreed level of throughput for a particular QCI.
The \texttt{throughputBreach} function would be triggered by the responsible monitoring component in the MNO; it specifies the culprit SCP, the QCI of interest, and the deficit throughput with respect to the agreed average. The contract will then apply a penalty in the form of a debit, here assumed to be proportional to the difference between the measured and the agreed throughput; this debit will be detracted from any existing or future credit to the SCP for payment for its services. We also included a simple "3-strike rule" in which, at the third consecutive infraction, the SCP is removed from the register, effectively preventing it from further transaction with the MNO. 

Naturally this contract is a proof of concept, with no claims of completeness or fitness for purpose; however, we believe that it  demonstrates how complex business agreements can easily be translated into automatically enforced reward and penalty mechanisms through the use of smart contracts. 
\vspace{-3pt}
\subsection{Smart Contract Limitations}
\label{ssec:limitations}
\vspace{-4pt}
While a detailed analysis of the issues related to the adoption of smart contracts is out of the scope of this paper, it is worth mentioning a few caveats that the interested parties should be aware of. The first issue is the legal validity of these agreements. As reported in \cite{fullbright_r3}, whether smart contracts can give rise to legally binding contractual relations depends on a number of factors, including the nature of the smart contract (i.e. whether they include or operate in conjunction with contractual terms), the specific jurisdiction in which the contract applies etc. Even where the smart contract has legally binding effects, enforcing them could be impossible due to the nature of the blockchain technology being used - e.g., there might be no central authority able to rectify a transaction over a permissionless ledger. 
Moreover, smart contracts are essentially pieces of code, and as such they are vulnerable to bugs and attacks. 
It is recommended to  include a fail-safe mechanism in the contract to be able to disable it and recover any outstanding balance in its possession in the event of the discovery of a vulnerability in the code.

\vspace{-4pt}
\section{Conclusions}
\vspace{-4pt}
In this paper we propose blockchain-supported Smart Contracts as an enabler to push SCaaS to individual users and retail venues. While advancements relating to multi-tenant small cells are needed  to allow small cell owners to aid network densification  and provide RAN capacity  in a local area, the cost and difficulty of establishing long-term business agreements with a plethora of individual partners represent a significant obstacle to pushing this sharing model to its logical extreme. Smart contracts have the potential to remove these obstacles by cheaply and effectively implementing simple business agreements -- in the form of SLAs -- between SCPs and MNOs. 

Regardless of whether the network of the future will use smart contracts like those we have presented here, we believe that using the disruptive power of blockchain technologies as a lens through which to re-think and re-invent standard practices of wireless communications will help us identifying critical areas that are ripe for innovation.
\vspace{-6pt}
\bibliographystyle{IEEEtran}
\bibliography{bibliography}

\end{document}